\documentclass[9pt,twocolumn,twoside]{osajnl}

\journal{jocn} 

\setboolean{shortarticle}{false}

\usepackage{algorithm}
\usepackage{stfloats}

\usepackage{acronym}
\acrodef{ML}{Machine Learning}
\acrodef{OPM}{Optical Performance Monitoring}
\acrodef{BAC}{Balanced Accuracy}
\acrodef{F1}{F1-score}
\acrodef{G-Mean}{Geometric Mean}
\acrodef{MLP}{Multilayer Perceptron}
\acrodef{FNR}{False Negative Rate}

\title{Data-driven mitigation of catastrophic forgetting in dynamic physical layer attack detection}

\author[1,*]{Aleksandra Knapi\'nska}
\author[2]{Marija Furdek}

\affil[1]{Department of Systems and Computer Networks, Wroc\l{}aw University of Science and Technology, Wroc\l{}aw, Poland}
\affil[2]{Department of Electrical Engineering, Chalmers University of Technology, Gothenburg, Sweden}

\affil[*]{aleksandra.knapinska@pwr.edu.pl}

\begin{abstract}
Optical networks are critical infrastructure that underpins global communications, and detecting security breaches that jeopardize them is  essential to maintaining worldwide connectivity. As malicious actors continuously evolve their attack techniques, dynamically updated intrusion detection models have become a key component of modern defense mechanisms. By incorporating newly acquired telemetry data, these models can adapt to emerging threats while maintaining high detection performance. However, when previously encountered attacks reappear after a prolonged period of absence, adaptive models may fail to recognize them due to the phenomenon of catastrophic forgetting. In contrast, statically trained models can reliably detect attacks represented in the original training data but lack the ability to adapt to previously unseen attack patterns. Consequently, intrusion detection systems face a fundamental tradeoff between adaptability to evolving threats and long-term retention of previously acquired knowledge. In this work, we propose a data-driven mechanism to cope with catastrophic forgetting in dynamic attack detection systems. Our approach balances the model update datasets by using parts of past attack data. We utilize a~threshold-based mechanism to trigger data balancing after accuracy drops due to an active attack change. Applied to an experimental optical network security dataset, the proposed approach reduces the average model adaptation time by 37\% compared to its dynamic counterpart that does not employ data balancing. Compared to a baseline from the literature that relies on neural network depth increasing, our approach requires 6\% fewer data batches to adapt to changing conditions and regain performance. 
\end{abstract}

\setboolean{displaycopyright}{false} 

\begin{document}

\maketitle

\section{Introduction}
Optical networks constitute the fundamental infrastructure supporting global communications. As a critical component, they are increasingly targeted by harmful activities and intrusions that may result with service disruptions or eavesdropping. 
Service degradation at the optical layer can be caused by, for example, targeted fiber cuts, insertion of harmful jamming signals, or polarization scrambling. Eavesdropping can be performed, for example, by fiber tapping via fiber bending or monitoring port access \cite{dahan2017security,uematsu2017design}.
Due to the extremely high data rates employed in optical networks, even short interruptions at the optical layer may result with large amount of data being lost, compromised or corrupt. Physical-layer breaches may also propagate to upper-layer services, potentially causing cascading and wide-spread service failures \cite{rak2021disaster}.

Maintaining service continuity in the presence of physical-layer breaches requires their timely and accurate detection, and tailored response. Real-time monitoring of the physical layer is paramount to its secure operation \cite{skorin2016physical}. In this context, security monitoring methods based on \ac{ML} are continuously developed to protect the network in an automated manner \cite{furdek2021optical}. 
Various \ac{ML} techniques have been demonstrated as paramount for optical network security management. They excel in identifying the related intricate and subtle effects in order to distinguish the causes of the known types of threats \cite{molina2020survey,rafique2017cognitive} as well as detect various novel physical-layer breaches \cite{natalino2026unified}. In multiclass scenarios, concrete attack types can be recognized based on the samples present in the dataset. On the contrary, in binary classification intrusions are distinguished from benign measurements. The latter approach also enables the recognition of novel attacks, as the model can find common patterns among acquired data from past attacks. 

\begin{figure*}[h]
    \centering
    \includegraphics[width=\linewidth]{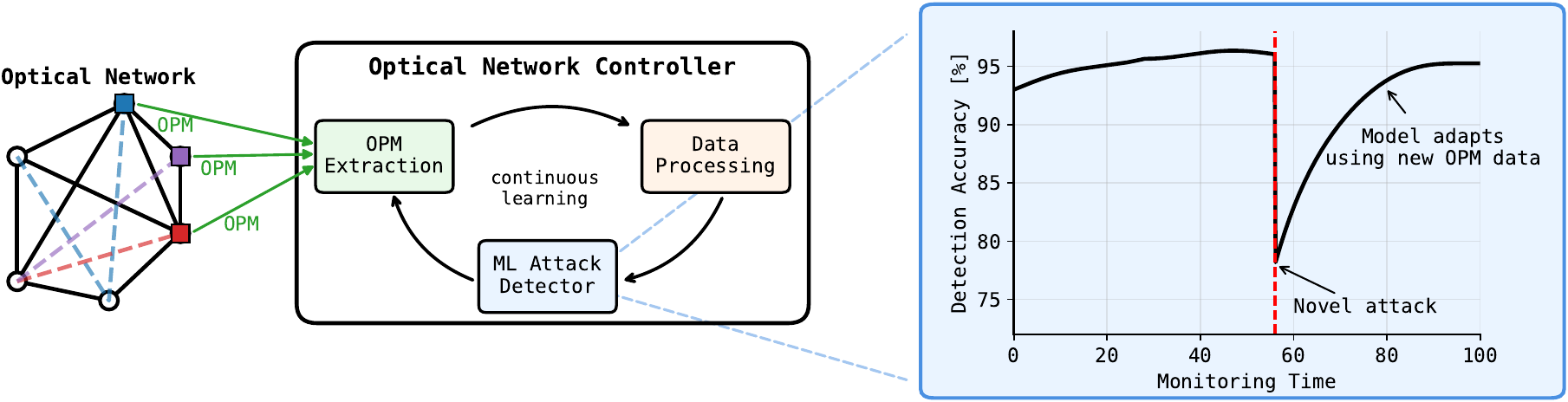}
    \caption{ML-based network monitoring system based on OPM telemetry and constantly updated security assessment based on acquired data. }
    \label{fig:overview_figure}
\end{figure*}

The complex and evolving network security landscape features new data transmission and control paradigms \cite{andriolli2022optical}, novel attack vectors, changing network attack surface \cite{ayodele2024sdn} and new types of attack techniques \cite{gazani2026optical}.
This requires continuous progress and development of new approaches that can cope with emerging network threats.
In our recent research \cite{knapinska2025experimental,knapinska2026dynamic}, we introduced methods to continuously integrate newly acquired knowledge about attacks into dynamic \ac{ML} models for physical-layer breach detection. 
We demonstrated the critical role of dynamic models in mitigating \emph{concept drifts}, which occur when the data distribution changes due to evolving network intrusions. These drifts can render the traditional, static models obsolete and result in misleading predictions. When an automated system remains outdated, unseen attack traffic may be mislabeled as benign, resulting in potential breaches leading to financial or reputational losses due to the absence of detection.  

Figure~\ref{fig:overview_figure} illustrates the high-level concept of optical network security monitoring system where the physical layer parameters are collected and analyzed to assess whether their values map to normal operating conditions, or if there may be an attack. An optical network controller featuring a dynamic, continuously learning \ac{ML}-based attack detector is able to consistently adapt and improve its quality. The \ac{OPM} parameters, indicating the health of optical services, are streamed by the optical transceivers into a data processing module that updates the attack detector with telemetry data from  previously encountered events.  As shown in the left half of the graph inset, the attack detector, once trained, usually achieves high accuracy in attack detection.
When a novel attack is introduced, as indicated with the red dashed line, the detection quality typically momentarily drops, as the related data differs from anything that was present in the training set to date. However, the continued feedback loop allows the system to adapt and regain the attack detection accuracy as it processes the newly-collected \ac{OPM} samples. 

While dynamic \ac{ML} models excel at learning about new attack types, compared to static ones, they may be more susceptible to \emph{catastrophic forgetting} of recurring attacks observed earlier. 
This effect is often observed in continual learning, where a detrimental performance drop can be observed for recurring, previously learned tasks \cite{wang2024comprehensive}. It gives rise to numerous challenges stemming from unexpected errors on tasks that were previously easily solved. For instance, a well-known attack that had been dormant for an extended period might cease to be recognized by the network controller’s \ac{ML} module, as it adapted to recognize more recent and sophisticated breaches. This highlights a trade-off between the adaptability and longevity of dynamic network intrusion detection models. A co-existing issue is the rarity of attack examples that can be quickly acquired, resulting in highly imbalanced network intrusion datasets. As the dynamic methods are designed to only process the new data, they are typically incapable of instantly regaining their accuracy, even if samples of a particular attack have previously been processed. 

To address this issue, in this paper, we propose a method for accelerating the readaptation of dynamic \ac{ML} models for network intrusion detection to previously encountered attacks. We focus on the binary scenario which allows for dynamic detection of previously unseen attacks. Building upon our preliminary study in \cite{knapinska2026dynamic}, we extend our framework by considering the recurring attacks scenario and developing specialized methods to address the challenge of catastrophic forgetting. While existing research focuses primarily on increasing the complexity of systems based on neural networks  \cite{kozal2023increasing,hayes2020remind,kirkpatrick2017overcoming} to cope with the increasing amount of data to be processed over the network lifetime, we take a novel, \emph{data-centric} view. Our proposed methodology uses historically acquired samples to gradually balance the training datasets used for continuous model updates. By carefully steering the update data imbalance ratio, our framework intelligently utilizes the samples acquired as the network was targeted by malicious actors. In particular, we balance the constant-size update data batches to provide the model with more attack samples and make the benign values no longer dominant. The resulting feedback loop, which is illustrated in Figure \ref{fig:overview_figure}, is curated to make the best use of the data that is collected over time. To this end, \ac{OPM} data from the previous occurrence of an attack are sampled among the collected data to improve the model adaptation in the current network state. This is done by replacing some of the recently collected benign \ac{OPM} samples, that would otherwise be used for model update, with the a bigger pool of available previously-encountered attack examples and keep the update data batch constant in size. 
To the best of our knowledge, this is the first study in physical-layer security that does not treat data imbalance as a constant descriptor, but selects the most effective portion of the available data for continuous improvement of the intrusion recognition quality.  

We evaluate the performance of our approach on an experimental optical networking dataset with six types of diverse physical layer intrusions \cite{furdek2020machine}. Compared to the standard dynamic setting which uses a constant portion of attack samples when an attack reappears, we decrease the model instability upon a reoccurrence of an attack, measured in standard deviation of \ac{BAC} between different attack types, by half. Compared to a baseline approach from the literature which relies on increasing the depth of the neural network model, our proposed approach achieves 6\% faster return to previous accuracy after an intrusion is reintroduced to the network. Moreover, while the baseline achieves faster model adaptation at an increase in model complexity coming from the additional neural network layer, the proposed method achieves the same performance without such overhead, relying on data curation instead. 
We further combine our \emph{data-based} methodology with the neural network depth-increasing baseline, and achieve an additional 37\% reduction in the adaptation time of the dynamic attack detector to recurring intrusions with improved stability manifested by a 73\% decrease in standard deviation of \ac{BAC} between attack sequence orderings. 

The remainder of this paper is organized as follows. Section~\ref{sec:related_work} presents the background concepts and reviews the related literature. Section~\ref{sec:proposed_model} presents the proposed solution based on balancing datasets using data acquired during evolving attacks. Section~\ref{sec:experimental_setup} presents the experimental scenarios and data considered in this work. Section~\ref{sec:results} presents a comparative performance analysis of the proposed approach and the baselines. Finally, Section~\ref{sec:conclusions} concludes this work. 

\section{Background and Related Work}
\label{sec:related_work}
This Section reviews prior literature to provide background on optical physical-layer security, data-driven intrusion detection, and methods for mitigating concept drift.

\subsection{Physical-Layer Vulnerabilities and Security Management}
Attacks exploiting vulnerabilities at the physical layer are typically categorized into those aimed at service disruption and eavesdropping. Service disruption techniques include in-band jamming, where a jamming signal overlaps with a targeted channel and adds unfilterable noise; out-of-band jamming, where a jamming signal is separated in spectrum and causes gain competition in erbium-doped fiber amplifiers; and polarization scrambling, where the fiber is squeezed at a high frequency, resulting in polarization modulation that causes transmission errors \cite{furdek2019experiment}. Eavesdropping can be performed by monitoring port access, evanescent coupling, V-groove cuts, and micro/macro bending \cite{harris1986bend,karlsson2022eavesdropping} that causes a portion of light to leak out from the fiber core and onto an eavesdropper's photodetector. 

To defend against these threats, physical-layer security management relies on three pillars: attack surface reduction, attack detection, and attack remediation. The first pillar encompasses understanding vulnerabilities, gauging the effects of attacks, and minimizing the attack surface through security-aware network design \cite{Skorin-Kapov:TON:2010,Hu:24,Li:2026}, service provisioning \cite{Zhu:JTL:2017, Savva:Access:2021}, and physical/optical layer encryption \cite{Liu:25}. The third pillar, devoted to attack remediation, encompasses techniques that neutralize the threat and recover affected optical services—for example, by rerouting to backup routes tailored to protect from breaches \cite{Furdek:JLT:2026}, or adapting frequency and modulation formats \cite{Papapavlou:2021}.

\subsection{Machine Learning for Optical Intrusion Detection}
Addressing the central security pillar--attack detection--prior studies have increasingly turned to data-driven approaches. Traditional intrusion detection systems rely on collecting telemetry data from field-deployed devices. In core optical networks, which are in the focus of this work, data is typically collected from commercial off-the-shelf coherent receivers that expose a rich dataset of \ac{OPM} parameters to the optical network controller \cite{furdek2020machine,natalino2022rootcause,natalino2022microservice}. Other types of parameters that can be used include the state of polarization of light \cite{sadighi2026ml,sadighi2025deep}.Specialized detectors based on supervised learning can be trained on this historical data to classify known attacks and raise alarms for future intrusions \cite{dasilva2024anomaly,kaur2023artificial,irram2022physical}, recognize multiple attack types \cite{sakhnini2021physical}, detect eavesdropping \cite{sadighi2024machine,sadighi2024detection}, or distinguish attack traffic from benign traffic \cite{hoang2021physical}. 

To address the limitations of supervised models against unseen attack types, prior work has explored unsupervised and dynamic learning techniques. Unsupervised learning approaches effectively identify anomalies in telemetry data, enabling operators to detect suspicious activities \cite{Natalino:JOCN:23,lechowicz2024trade} and analyze their root causes \cite{natalino2022rootcause}. Furthermore, continuously updated binary classifiers trained on datasets with multiple attack types have been proposed to allow models to adapt and improve over time \cite{knapinska2025dynamic,knapinska2025experimental}.

\subsection{Concept Drift and Adaptation Strategies in Intrusion Detection}
When network intrusions evolve over time, changes in the underlying data distribution occur--a challenge known in \ac{ML} as \emph{concept drift} \cite{lu2019learning}. To maintain high detection performance under concept drift \cite{aguiar2024survey}, several adaptation strategies have been investigated in prior literature. Ensemble learning continuously updates sets of classifiers to balance legacy knowledge with new data \cite{krawczyk2017ensemble}. Window-based retraining methods periodically retrain models from scratch as new data batches arrive \cite{yu2025adaptive}. More advanced approaches update neural networks incrementally through \emph{partial fitting} \cite{knapinska2025dynamic,andresini2021network}, fine-tuning models like \acp{MLP} without resetting their weights. Self-supervised learning generates pseudo-labels for newly observed data when annotations are sparse \cite{yang2025self}. Finally, continual learning methods adapt to evolving attacker techniques \cite{zhang2025continual,fuhrman2025cnd} and enhance model robustness \cite{kozal2023defending} by enabling learning of successive tasks while retaining past knowledge.

\subsection{Recurring Concept Drift and Research Gap}
A particular type of concept drift is \emph{recurring concept drift}, in which previously observed concepts reappear after a period of absence \cite{gama2014survey}. This phenomenon naturally occurs in network security monitoring when previously encountered attacks reemerge. As an example, consider an optical communication system operating over three consecutive time periods. During the first period, the network experiences a polarization scrambling attack, and the intrusion detection model learns to recognize its characteristic signatures from \ac{OPM} data. In the second period, the attacker switches to a jamming attack, previously unknown to the model, requiring it to adapt to a different distribution of observations. Finally, the attacker returns to the original polarization scrambling attack. A dynamic model that has completely adapted to the jamming attack may no longer recognize the signatures of polarization scrambling, even though it had previously learned them. This phenomenon, in which a model loses previously acquired knowledge while learning new concepts, is known as \emph{catastrophic forgetting} \cite{french1999catastrophic}. Conversely, relying solely on a static model is also insufficient, as it cannot adapt to genuinely new attack patterns absent from the training data. Consequently, adaptive intrusion detection models must balance learning new concepts while retaining previously acquired knowledge that may become relevant again when attacks reoccur. 

To the best of our knowledge, the issue of reoccurring network attacks was only briefly discussed in the context of the IP layer through a specialized neural network architecture \cite{ying2024metanoia}. However, it has never been addressed in the context of optical network security. This work closes this research gap by proposing a new approach based on utilizing telemetry data that is collected by the network operators by default, without requiring specialized monitoring devices.

\section{Data-driven model updating}
\label{sec:proposed_model}

The approach proposed in this work is based on the concept of utilizing continuously collected data to balance updating data for dynamic attack detectors. 
Typical network conditions provide abundance of data  representing normal operating state, and scarce attack samples. 
Recent studies in meta-learning, which exploits dataset meta-features to characterize learning problems and support algorithm selection, indicate that dataset imbalance is one of the key factors contributing to the difficulty of classification tasks \cite{lorena2019complex,komorniczak2023problexity}. Consequently, balancing the training datasets can simplify the classification task. Building upon this observation, our proposed framework utilizes the past attack data collected during network operation. Contrary to conventional dynamic methods which only process the newly incoming data, we make use of historical samples that are anyway collected and stored by the operators. Thus, the proposed method enables reusing valuable past observations that would not have been re-accessed if a traditional dynamic method was used. 

\begin{figure}[t]
    \centering
    \includegraphics[width=\linewidth]{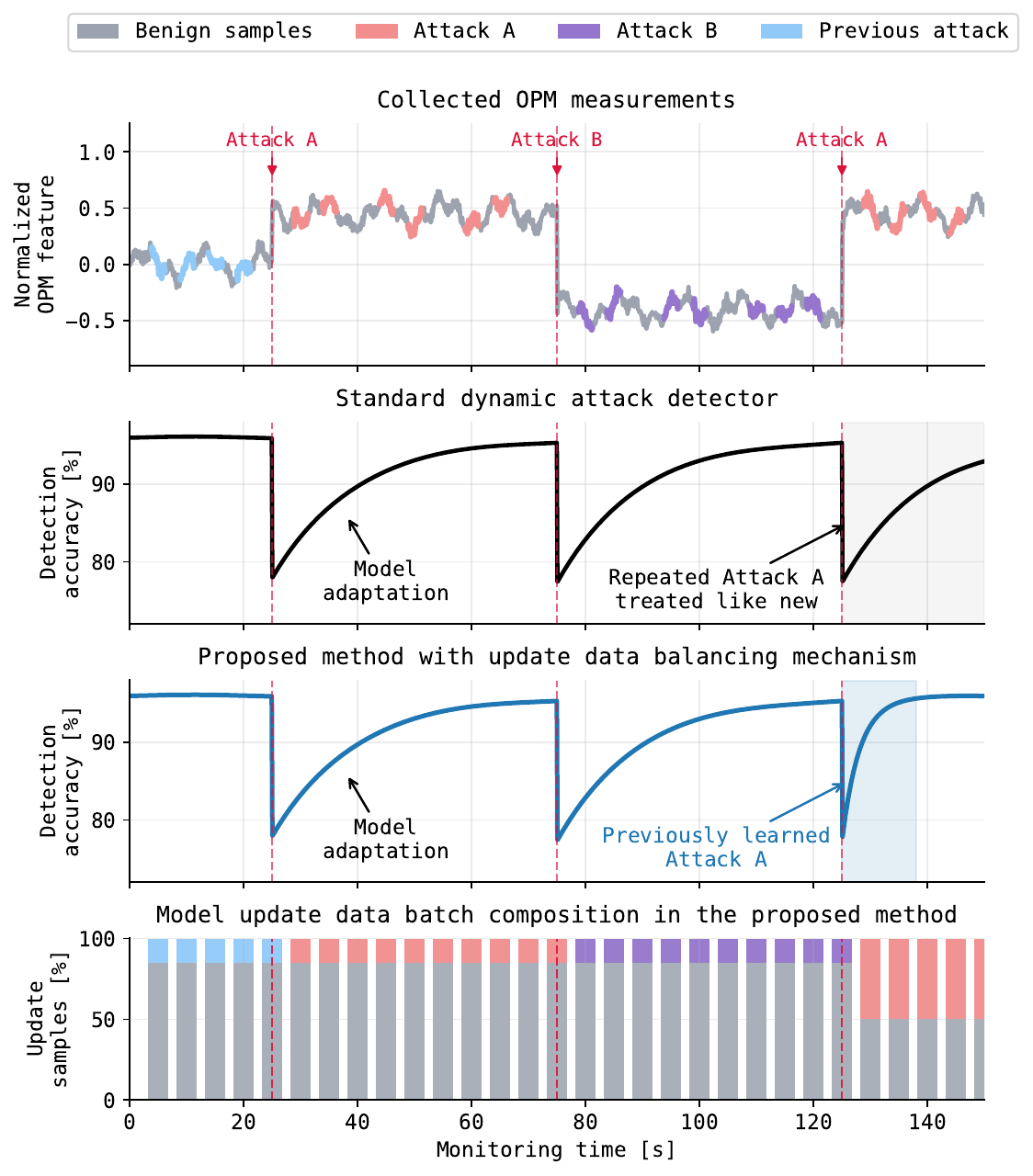}
    \caption{The proposed framework of updating the attack detection model with balanced datasets after an attack reappears.}
    \label{fig:method_illustration}
\end{figure}

Figure~\ref{fig:method_illustration} illustrates the underlying idea of the methodology. As more attack data becomes available over time, the dataset used to update the \ac{ML} model can gradually become more balanced, enhancing the model's performance. Let us consider the detailed sequence of actions shown in the figure.

Initially, access to attack samples is limited, and benign samples, depicted with gray color, constitute the majority of the training data. In the figure, around 25 s, attack type A is inflicted and the model performance plunges. To adapt, the dynamic model is trained and updated using an imbalanced dataset, as new attack samples are extremely rare. Our previous work \cite{knapinska2025experimental,knapinska2026dynamic} demonstrated that gradually updating the detector with samples of the current intrusion in such an imbalanced setting improves detection accuracy. 
This improvement is achieved without any architectural modifications to the model. It is also quick, as the model update on a batch of 100 samples acquired per minute takes under 6 ms. 
An attacker can bypass the benefits of dynamic model adaptation by changing the attack type, denoted as introducing Attack B in the figure, which triggers another cycle of performance drop followed by adaptation. The second graph from the top shows the performance of a standard dynamic attack detector. When the attacker re-introduces Attack A, the model, affected by catastrophic forgetting, again treats it as new attack and initiates the same adaptation process. 

Our proposed approach, illustrated in the third graph from the top, addresses the issue of catastrophic forgetting by balancing the dataset used for model update with historical attack data instead of waiting for scarce new samples to accumulate. To identify whether historical data that represents this attack type is available in the telemetry database and should be pulled for model update, nearest-neighbor matching with a distance threshold or a similar approach can be used. 
To maintain a constant batch size, benign examples are then randomly sampled from the available pool, and attack examples are retrieved from the historical data. This approach creates a mixture of current and past samples from this attack type, resulting in a balanced dataset that improves accuracy more rapidly. This process is also referred to as the \emph{rehearsal strategy} \cite{maltoni2019continuous,benkHo2024example}, where past information is periodically replayed to the model to strengthen connections for memories it has already learned. Thus, a network operator can effectively utilize the data acquired over time under various conditions, with the expected significant improvement in performance without any architectural modifications.

\begin{algorithm}[t]
\footnotesize
\caption{Balancing the update dataset to overcome catastrophic forgetting.}
\label{alg:balancing_strategy}
\begin{algorithmic}[1]
    \State \textbf{Input:} Buffer size $S$, accuracy threshold $T$
    \While{OPM samples are incoming} \label{line:samples_incoming}
        \For{each sample}
            \State classify the sample as attack or benign \label{line:classify_sample}
            \State save the sample in the historical data buffer
            \If{historical data buffer contains $S$ samples} \label{line:buffer_full_check}
                \State calculate model accuracy on the samples in the buffer \label{line:test}
                \If{model accuracy $< T$}
                    \State N = number of attack samples in the buffer \label{line:count_current_attack_samples}
                    \State calculate the number of additional attack samples required 
                    \label{line:determine_number_of_needed_samples}
                    \Statex \hspace{\algorithmicindent}\hspace{\algorithmicindent}\hspace{\algorithmicindent}\hspace{\algorithmicindent}for a balanced class distribution as $M = \frac{S}{2} - N$
                    \State randomly select $M$ benign samples from the buffer \label{line:random_select}
                    \State replace the selected benign samples with $M$ attack samples \label{line:replace_samples}
                    \Statex \hspace{\algorithmicindent}\hspace{\algorithmicindent}\hspace{\algorithmicindent}\hspace{\algorithmicindent}from the telemetry database 
                \EndIf
                \State update the model using the samples in the buffer
                \State move the samples from the buffer to the telemetry database
                \State empty the buffer
            \EndIf
        \EndFor
    \EndWhile
\end{algorithmic}
\end{algorithm}

The proposed method is formally presented in Algorithm \ref{alg:balancing_strategy}. The base \ac{ML} model is a dynamic attack detector that is periodically updated with continuously acquired \ac{OPM} data (line~\ref{line:samples_incoming}). Being a crucial a part of the real-time network monitoring system, the model scrutinizes every incoming telemetry sample (line~\ref{line:classify_sample}). As a dynamic detector operating in a \emph{data stream mining} setting, it employs \emph{chunk-based} learning. In this setting, the model is updated after a predefined number of new samples $S$ (e.g., 100) has been collected (line \ref{line:buffer_full_check}). We utilize the \emph{test-then-train} protocol, where the model performance on the buffer is evaluated 
(line \ref{line:test}) and forms the basis for the decision stage. We assume that the labels become available for the samples waiting in the buffer, as the operator can retroactively mark them in case an undetected intrusion took place. In case a novel attack type has been introduced, the accuracy drops sharply. Thus, a~pre-set accuracy threshold $T$ (e.g., 90\%) is considered as a deciding factor. Model underperformance triggers the balancing strategy. To this end, the attack samples are counted (line \ref{line:count_current_attack_samples}). With the fixed buffer size, it’s straightforward to calculate the required number of samples to achieve a~balanced dataset, as they should represent half of the total (line~\ref{line:determine_number_of_needed_samples}). The corresponding number of benign samples from the current data batch is then randomly chosen (line \ref{line:random_select}) and replaced with attack samples from the telemetry database (line~\ref{line:replace_samples}). 
The resulting \emph{data batch} is then used to \emph{partially fit} the model. If the accuracy was above the selected threshold, then the original, imbalanced batch is used. The specific model update procedure depends on the underlying \ac{ML} algorithm. For a neural network, a single optimization pass is performed to update the model weights. For an ensemble model, the update may instead involve a pruning procedure. 
The proposed methodology is model agnostic, and compatible with any \ac{ML} algorithm capable of incremental learning, or with a classifier ensemble.

\section{Experimental setup and dataset}
\label{sec:experimental_setup}
\begin{figure}[h]
    \centering
    \includegraphics[width=\linewidth]{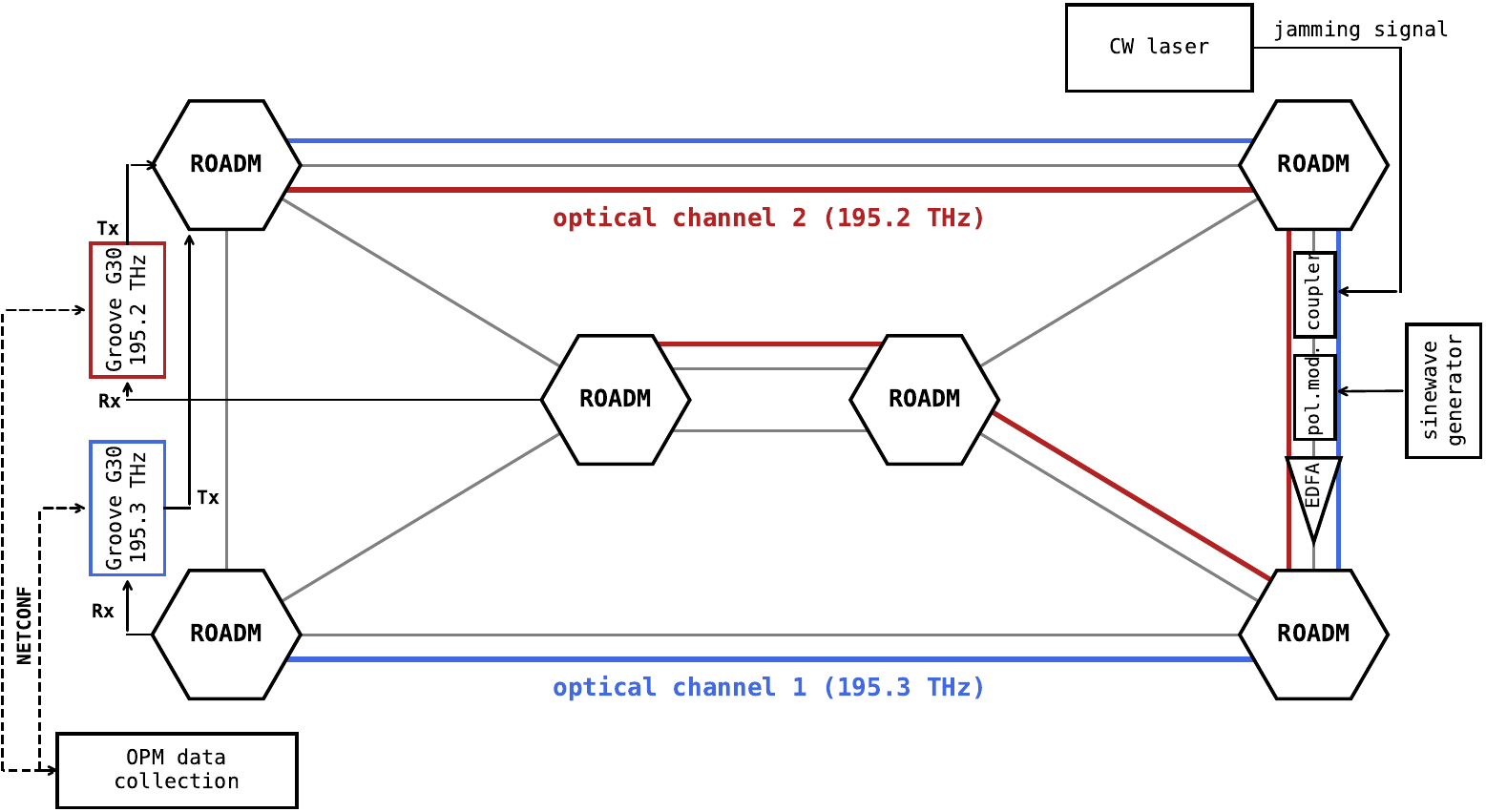}
    \caption{Experimental testbed where the dataset of \ac{OPM} para\-me\-ters from attack scenarios was collected \cite{furdek2019experiment,furdek2020machine}. }
    \label{fig:experimental_setup}
\end{figure}

The dataset used in this work consists of \ac{OPM} samples collected during a range of attacks performed in a laboratory environment described in \cite{furdek2019experiment}. The physical-layer security data analyzed in this study was gathered through a series of experiments in which an optical network testbed was exposed to multiple attacks designed to disrupt service operation. The testbed includes six nodes equipped with coherent transceivers and Reconfigurable Optical Add-Drop Multiplexers (ROADMs), one node containing an Erbium-Doped Fiber Amplifier (EDFA), and 10 optical fiber links. The two monitored channels correspond to two optical 200 Gbit/s polarization-multiplexed 16-Quadrature Amplitude Modulation (16QAM) signals centered at 195.2 and 195.3 THz. An overview of the experimental setup used for data collection is shown in Figure \ref{fig:experimental_setup}. A comprehensive description of the testbed and the conducted experiments can be found in previous work \cite{furdek2019experiment,furdek2020machine}.

Each attack type, as well as normal traffic, is represented by 1440 datapoints corresponding to minute-by-minute one-day traces, described using 31 distinct \ac{OPM} parameters summarized in Table \ref{tab:opm-parameters}. It is important to note that the parameters used in the dataset are those already monitored and recorded by network operators. This ensures that the detectors developed and analyzed in this work can be readily deployed in real-world environments without requiring additional instrumentation or data collection. 

\begin{table}[h]
    \centering
    \caption{\ac{OPM} parameters of each data sample.}
    \begin{tabular}{p{5cm} p{3cm}}
        \toprule
        \bfseries Description & \bfseries  Acronym \\
        \midrule
        Chromatic Dispersion & CD \\
        Differential Group Delay & DGD\\
        Optical Signal to Noise Ratio & OSNR\\
        Polarization Dependent & Loss PDL\\
        Q factor & Q-factor\\
        Bit Error Rate before FEC & BER-FEC\\
        Bit Error Rate after FEC & BER-POST-FEC\\
        Optical Power Received & OPR\\
        Optical Power Transmitted & OPT\\
        Optical Frequency Received & OFR\\
        Loss Of Signal & LOS \\
        \bottomrule
        \multicolumn{2}{p{8.5cm}}{\footnotesize The maximum, minimum and average values per 1-min observation interval are reported for all parameters except LOS.
        }
    \end{tabular}
    \label{tab:opm-parameters}
\end{table}

We consider three types of attacks: in-band jamming (INB), out-of-band jamming (OOB), and polarization scrambling (POL).
For in-band jamming, a low-power jamming signal is inserted within the bandwidth of the signal under test through a passive coupler, introducing unfilterable noise. We consider two power levels for the jamming signal: 10 and 7 dB below the channel under test, corresponding to a lighter (INBLGT) and a stronger (INBSTR) attack intensity, respectively.

Out-of-band jamming employs an intrusion signal with a frequency separate from the channel’s bandwidth and higher power, resulting in a reduction in the amplifier gain allocated to the legitimate channel. We also consider two power levels for the jamming signal: 3 and 8.7 dB above the channel under test, representing a lighter (OOBLGT) and a stronger (OOBSTR) attack intensity, respectively.

Polarization modulation attack involves attaching a polarization modulator to the fiber, comprising a fiber squeezer that induces stress-induced birefringence and causes polarization modulation. The induced changes occur at a faster rate than the coherent receiver’s polarization recovery algorithm, leading to burst errors without affecting the power or frequency of the transmitted signal. The modulator is driven by a sine wave signal with a frequency of 136 kHz and peak-to-peak amplitudes of 0.4 and 1.6 V, respectively, simulating a lighter (POLLGT) and a stronger (POLSTR) attack intensity.  

\begin{figure}[h]
    \centering
    \includegraphics[width=.6\linewidth]{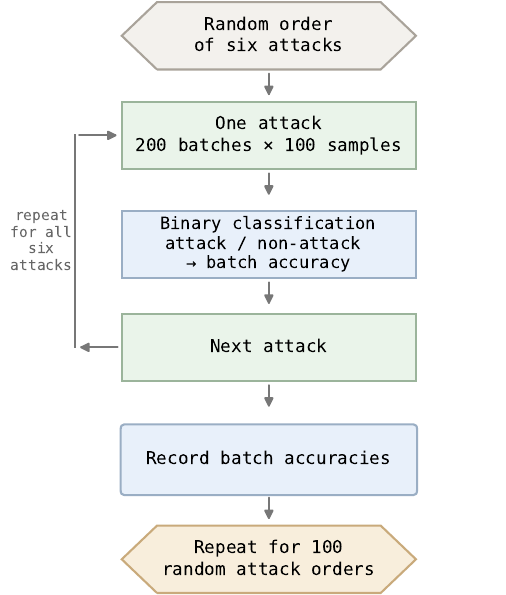}
    \caption{Illustration of the experimental scenario--streaming an attack sequence for binary classification. }
    \label{fig:scenario_illustration}
\end{figure}

In the \ac{ML} experiments conducted in this paper, we initially employ a 15:100 attack-to-normal ratio. Each experiment is replicated 100 times, resulting in 100 random stratified data samplings from the original, balanced dataset \cite{furdek2020machine}. As we consider a streaming scenario, the six attacks are introduced to the binary detection system in a sequential manner. To achieve this, while maintaining the 15:100 class distribution, we inject attack samples from only one type at a time. Specifically, the first 200 batches contain a single attack type, the subsequent 200 batches a different one, and so forth. All attack types share the same label, ensuring that the model remains unaware of the changing nature of the attacks over time. This scenario is illustrated in Figure \ref{fig:scenario_illustration}. Given that our dataset encompasses six attack types, there are 720 possible attack orderings. In the subsequent section, we analyze the results averaged over 100 randomly selected sequences. 

\begin{figure*}[htb]
    \centering
    \includegraphics[width=\linewidth]{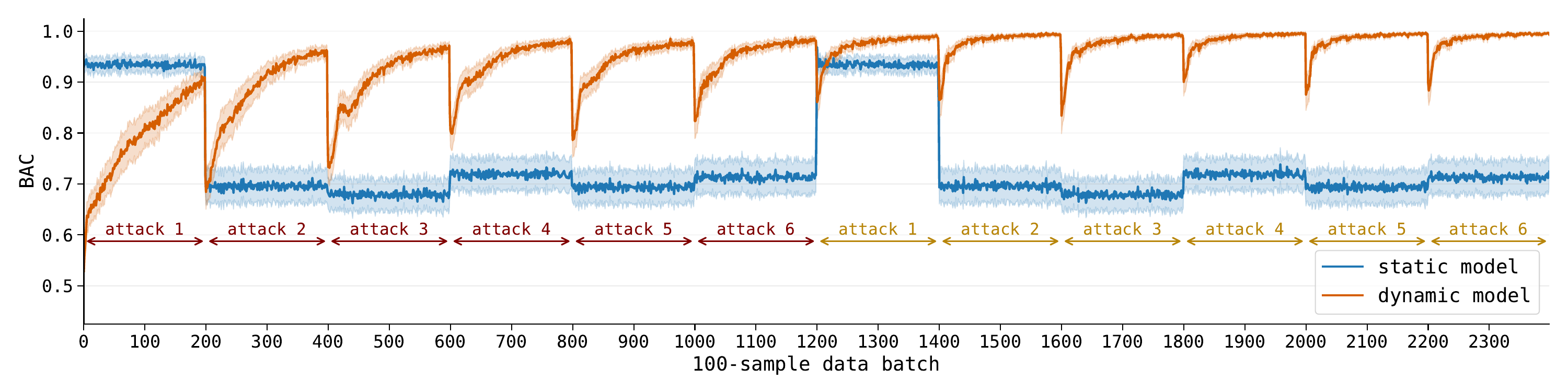}
    \caption{Static vs dynamic model with recurring attacks. Average BAC from 100 experiment replications (each replication consisted of a random attack sequence repeated twice) with confidence intervals. }
    \label{fig:static_dynamic_recurring}
\end{figure*}

\begin{figure*}[h]
    \centering
    \includegraphics[width=\linewidth]{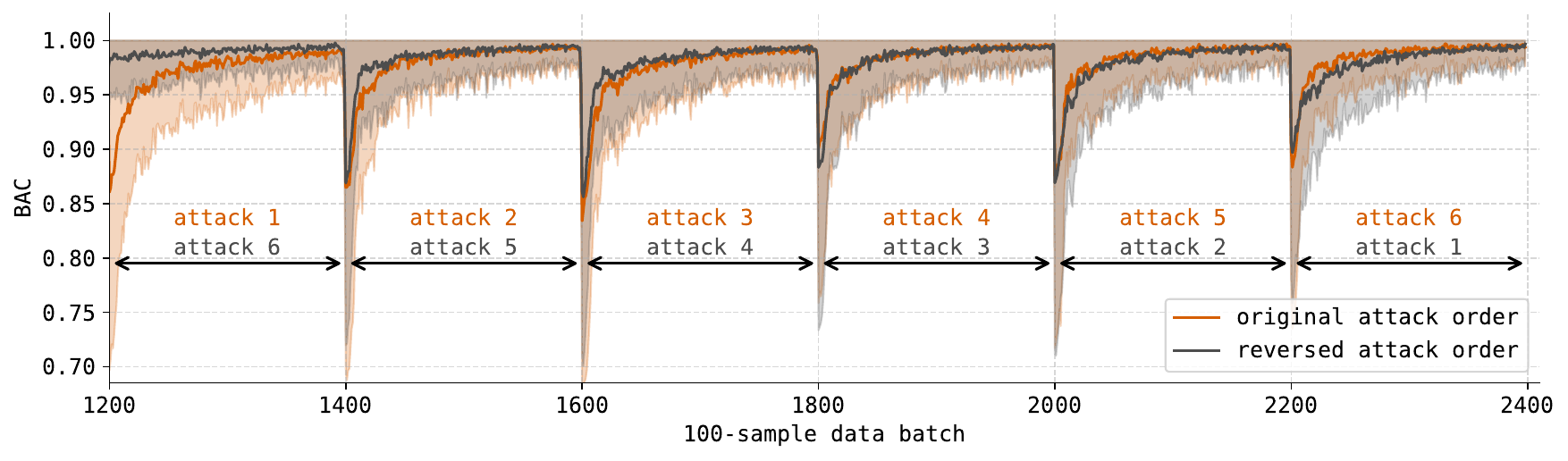}
    \caption{Imbalanced original vs reversed attack order recurring attacks (only the recurring part--see x-axis). Average BAC with standard deviation from 100 experiment replications (attack orderings). }
    \label{fig:original_vs_reversed}
\end{figure*} 

To enable a direct comparison with literature baselines, we utilize the \ac{MLP} as the base \ac{ML} algorithm. We employ the architecture that proved effective for this task in previous work \cite{knapinska2025experimental,knapinska2026dynamic}--a single hidden layer of 100 neurons. To update the model with subsequent data batches, we use the \emph{partial fit} procedure \cite{pedregosa2011scikit}, which performs one optimization pass on each provided mini-batch while preserving the learned model parameters between calls. This allows the network to be updated iteratively as new data become available without reinitializing the model. In the experiments reported in \cite{knapinska2025experimental}, the periodical model update takes approximately 5 ms on a laptop if this \ac{MLP} architecture is used with 100-sample batches. 

A baseline solution from the literature considered in this work is a recently proposed idea to increase the depth of the neural network employed for attack detection \cite{kozal2023increasing}. This method is effective as the layers responsible for past tasks are sustained \cite{hayes2020remind,kirkpatrick2017overcoming}. To compare the proposed scheme with the baseline, we begin the experimental part with the same architecture of a single hidden layer containing 100 neurons. In the increased depth modification, we introduce a second hidden layer with 50 neurons. We evaluate two versions of this architecture--with and without freezing the first hidden layer before adding the second one. 

\section{Results}
\label{sec:results}
To assess the impact of attack reappearance on detection accuracy, we first conduct exploratory experiments to verify its effects on both traditional static and dynamic models. After confirming the presence of catastrophic forgetting and establishing the need for mechanisms to increase the model’s adaptation speed, we compare the performance of the proposed methodology with the literature baselines. We evaluate performance in terms of the average \ac{BAC} obtained for each batch of 100 samples. We also evaluate the time required for the model to adapt to a change in the active attack. Although the proposed method can be applied to various \ac{ML} models and batch sizes, to ensure a fair comparison between the methods, we define adaptation time as the number of data batches that must be processed to achieve a \ac{BAC} of 0.95 following a change in the active attack. 

\subsection{Impact of attack reappearance on traditional static and dynamic detectors}

Figure \ref{fig:static_dynamic_recurring} shows the performance of the static and dynamically-adapting \ac{MLP} models as in \cite{knapinska2026dynamic}. Both models were trained on imbalanced attack data without any additional mechanisms to cope with catastrophic forgetting. The experiment demonstrates the expected behavior in terms of \ac{BAC}. The statically trained neural network is able to correctly detect attack types seen during training, even if they reappear (which is the case for attack 1). However, it is not able to adapt to unseen types of intrusions (all other attack types). In contrast, the dynamically adapting model is improving its detection accuracy as it acquires more data about the current network state. The sudden drop in detection accuracy is smaller with every new attack that is introduced to the network. However, the reappearance of previously encountered attacks still causes a drop, highlighting a tradeoff between adaptability and longevity of dynamic network intrusion detection models.

\begin{figure*}[h]
    \centering
    \includegraphics[width=\linewidth]{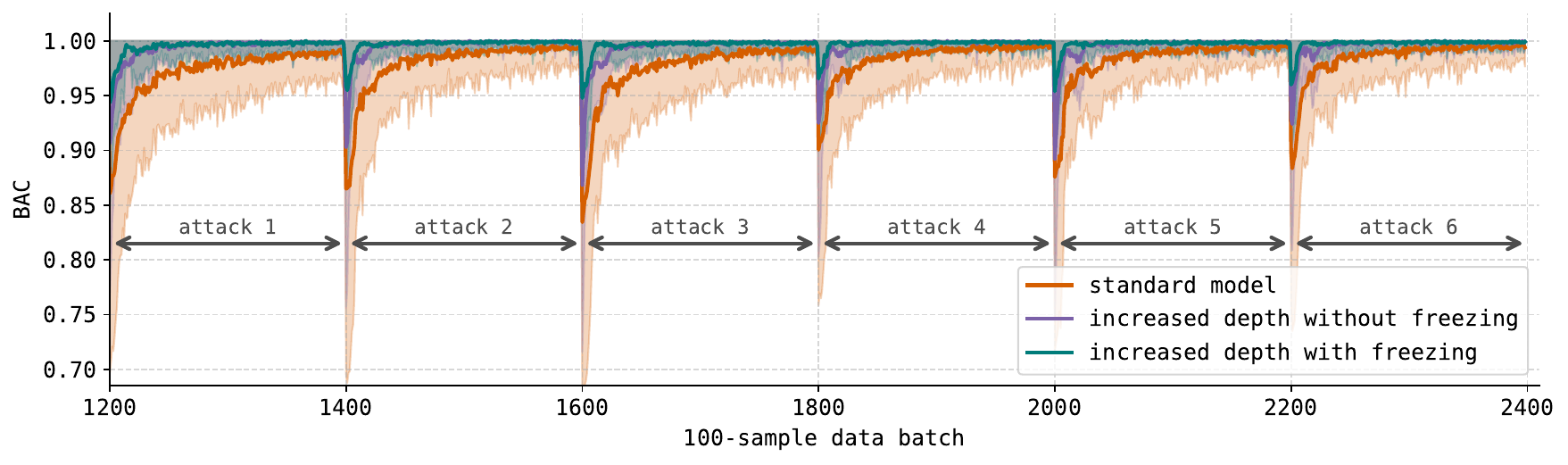}
    \caption{Standard model vs increased depth model with and without freezing hidden layers for recurring attacks (only the recurring part--see x-axis). Average BAC with standard deviation. }
    \label{fig:increased_depth}
\end{figure*} 

\begin{figure*}[htb!]
    \centering
    \includegraphics[width=\linewidth]{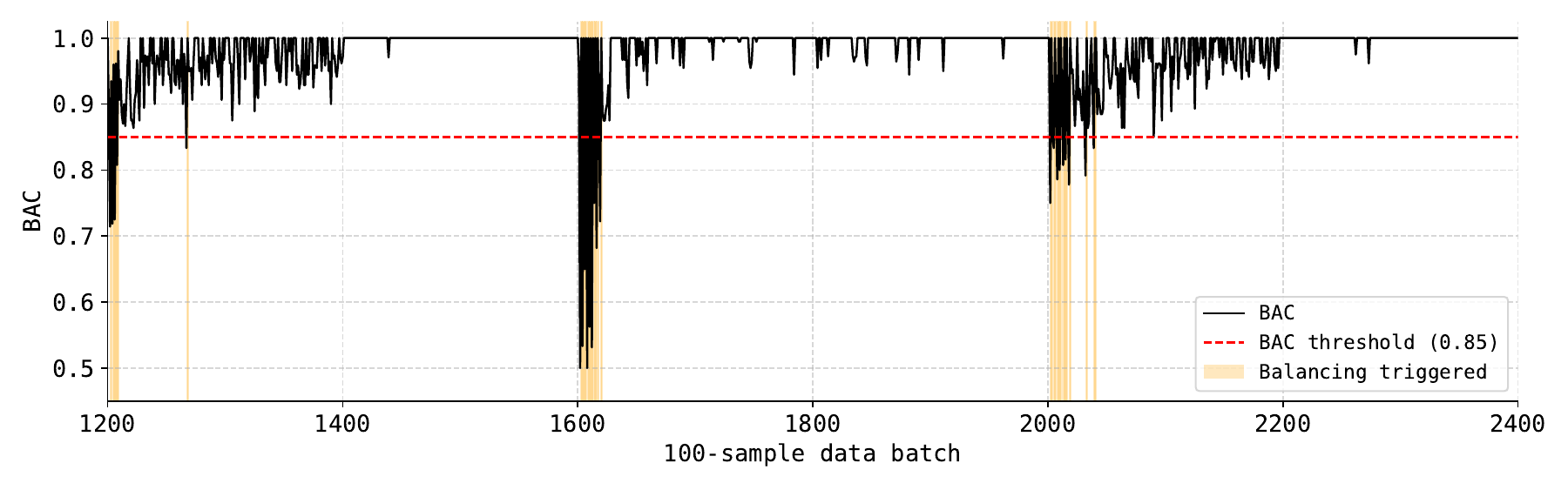}
    \caption{BAC over time for an example experiment run (attack order) with threshold 0.85. }
    \label{fig:threshold_0_85}
    
    \includegraphics[width=\linewidth]{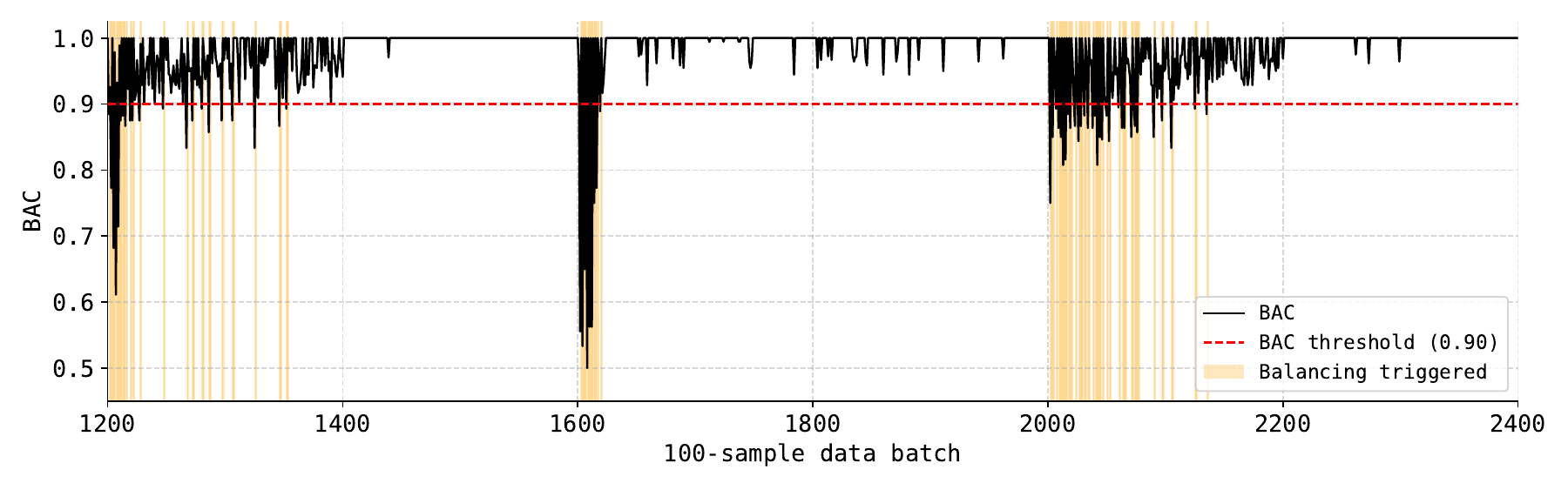}
    \caption{BAC over time for an example experiment run (attack order) with threshold 0.90. }
    \label{fig:threshold_0_90}

    \includegraphics[width=\linewidth]{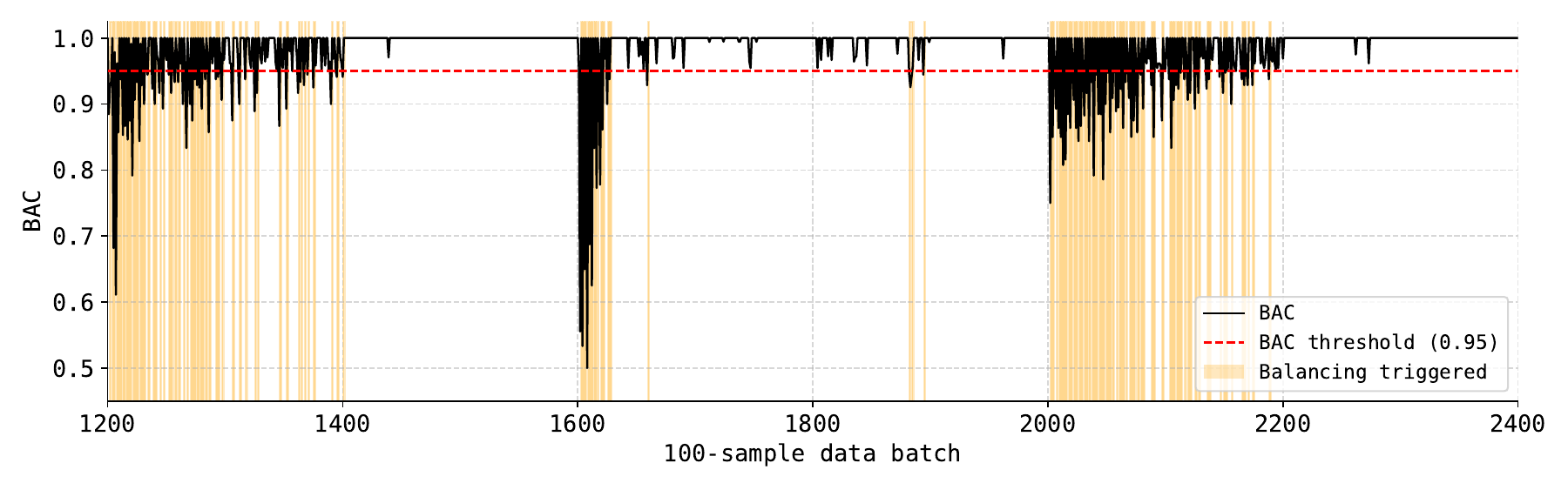}
    \caption{BAC over time for an example experiment run (attack order) with threshold 0.95. }
    \label{fig:threshold_0_95}
\end{figure*}

Based on the discussed results, the remainder of the experimental part focuses on the dynamic models. The second experiment evaluates the impact of the attacks' order of (re)appearance. To this end, we placed samples of the six attacks considered in this work in a sequence and then repeated the experiment in the same and the reversed order. The results, depicted in Figure \ref{fig:original_vs_reversed}, indicate that the order does not have a discernible impact on the detector’s accuracy when no additional mechanisms are introduced. The figure presents a zoomed-in section of the results, focusing on the second segment (attack reappearance), as indicated by the x-axis. The leftmost section of this segment presents the most intriguing insight: the active attack in the "reversed" case was not altered there, i.e,. it is the same as the attack in the preceding section (not shown). Consequently, there is no decline in accuracy. Nevertheless, there are almost no differences in the average and standard deviation of BAC in the subsequent attacks introduced in the experiment. The catastrophic forgetting is, thus, present regardless of how much time passed from the past encounter of a particular attack. 

The final experiment in this subsection assesses the efficacy of the baseline solution in the context of attack detection. We investigate whether the solution proposed in the literature, namely the increase of the depth of a neural network, is effective in the dynamic attack detection scenario. 
Figure \ref{fig:increased_depth} presents the results, depicting the \ac{BAC} from the second occurrence of the attacks (note that the x-axis starts at data batch 1200). Indeed, the average \ac{BAC} is higher for the increased-depth \ac{MLP}, with and without freezing hidden layers, demonstrating the solution’s effectiveness in the physical layer security use case. In each of the segments, the mean \ac{BAC} of the increased depth model is on average 2\% higher. Furthermore, the difference in standard deviation across the 100 attack orderings is as high as 165\%, illustrating the model stability regardless of attack sequence. Consequently, the model of increased depth performs better on average and exhibits significantly greater stability, irrespective of the attack reappearance order. This demonstrates its effectiveness as a method of mitigating catastrophic forgetting. 

\subsection{Performance evaluation of the proposed mechanism}

The aim of the first experiment in this part is to tune our proposed approach. To this end, we investigated the impact of the threshold on accuracy and balancing frequency by testing three \ac{BAC} threshold values: 0.85, 0.90, and 0.95. The results of an example experiment run are shown in Figures \ref{fig:threshold_0_85}, \ref{fig:threshold_0_90}, and \ref{fig:threshold_0_95}, corresponding to the respective threshold values. As expected, setting a higher threshold triggers more frequent balancing events, particularly after a change in the active attack, when the neural network model is adapting to the new conditions. However, a higher threshold does not necessarily translate into higher overall detection quality. The remaining experiment replications, conducted with different attack orders, yielded consistent results. For the considered dataset, a threshold of 0.90 provided a good balance between balancing frequency and model adaptation speed. After a change in the active attack, the average time required to regain \ac{BAC} of 0.95 required on average 0.42 batches less when the threshold was set to 0.95, compared to the 0.90 threshold. In contrast, the threshold of 0.85 required, on average, 5.5 additional batches to regain \ac{BAC} of 0.95. 

\begin{figure*}[h]
    \centering
    \includegraphics[width=\linewidth]{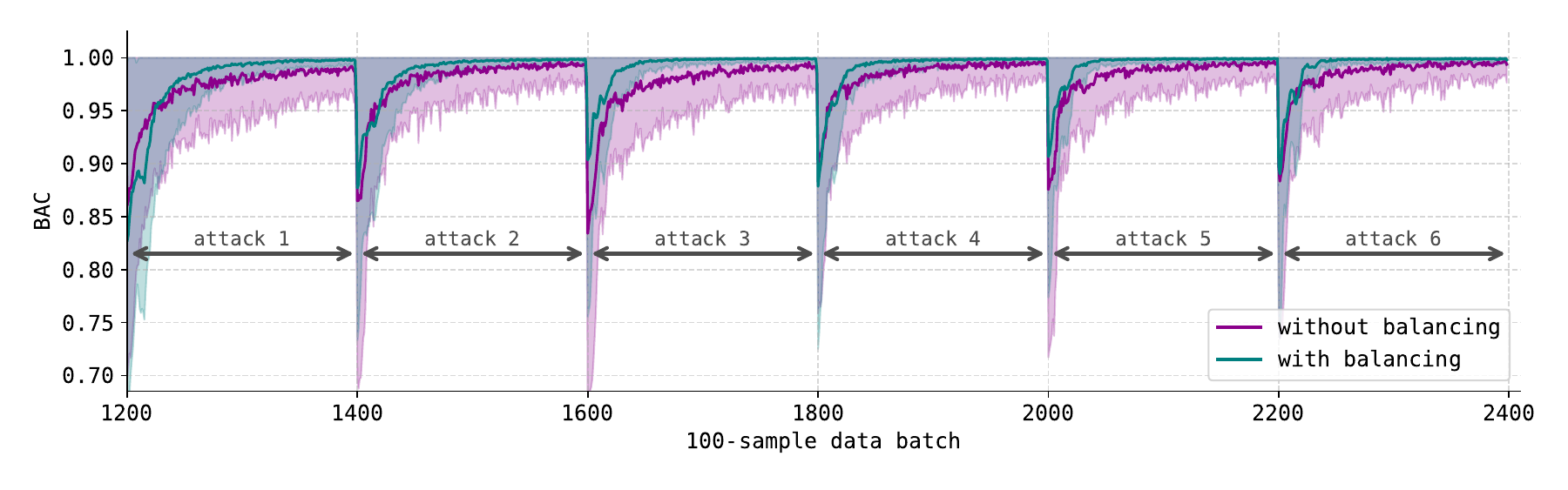}
    \caption{Imbalanced vs imbalanced to balanced with recurring attacks (only the recurring part--see x-axis). Average BAC with standard deviation. }
    \label{fig:imbalanced_vs_imbalanced_to_balanced_recurring}
\end{figure*}

Let us now evaluate the performance of the proposed solution, which doesn’t require any model modifications but rather changes how data is used, against baselines. 
Figure \ref{fig:imbalanced_vs_imbalanced_to_balanced_recurring} shows the benefits of balancing the training dataset of known attacks over time. As for the previous experiments, the plot shows the second occurrence of the attacks, as indicated on the x-axis. Despite not modifying the model, the method proves effective. The average \ac{BAC} improves slightly within each attack section and is on average 1\% higher than for the case without balancing. More importantly, the standard deviation is significantly reduced--on average it is 110\% lower when the balancing mechanism is introduced. With 100 experiments repetitions, that means that the results are more stable irrespective of the random attack sequence. This effect is shared with the literature benchmark of increasing the neural network depth. Another result that is improved is the convergence speed after the change of attack conditions. The balancing mechanism allows achieving the mean accuracy in each segment on average 13.5 batches faster than the standard dynamic approach using imbalanced data. In other words, upon an attack type change, the recognition quality is regained after processing on average 1350 samples fewer. 

\begin{figure*}[t]
    \centering
    \includegraphics[width=\linewidth]{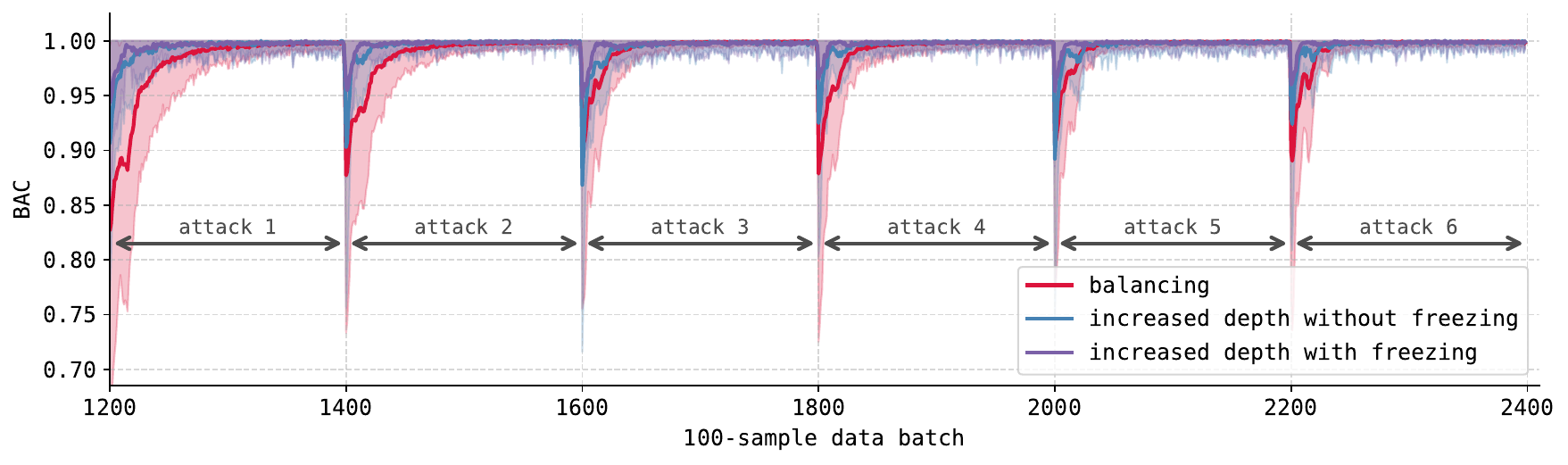}
    \caption{Balancing the dataset vs increasing depth of the neural network with recurring attacks (only the recurring part--see x-axis). Average BAC with standard deviation. }
    \label{fig:balanced_vs_increased_depth_recurring}
\end{figure*}

Figure~\ref{fig:balanced_vs_increased_depth_recurring} compares the performance of our approach to both versions of the literature benchmark. The results show that the proposed mechanism of balancing the dataset without changing the model architecture achieves performance similar to the benchmark that relies on increasing the model complexity. The main difference is in their convergence speed. Specifically, here we measured the time it takes to reach the median detection accuracy after the active attack is introduced. The increased neural network depth approach takes an average of 127 data batches to achieve this, while the data balancing approach requires 119 data batches--which is 6.3\% less. On the other hand, the time to regain an acceptable \ac{BAC} of 0.95 is faster for the literature method of increasing depth by 13 data batches on average. It is also important to note that the biggest performance difference is visible in the first two attacks. It suggests that the dynamic models successfully learn the baseline better as time passes. However, the difference becomes much smaller as more data is processed. 

\begin{figure*}[t]
    \centering
    \includegraphics[width=\linewidth]{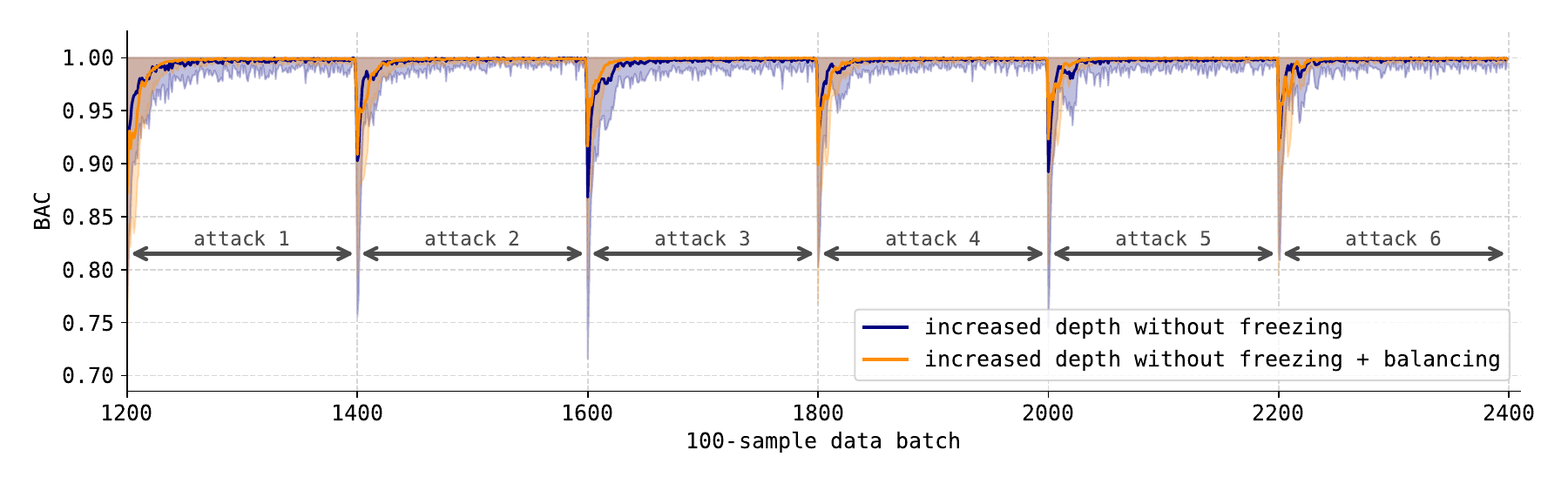}
    \caption{Balancing the dataset and increasing depth of the neural network with recurring attacks (only the recurring part--see x-axis). Average BAC with standard deviation. }
    \label{fig:balanced_and_increased_depth_recurring}
\end{figure*}

\begin{figure*}[t]
    \centering
    \includegraphics[width=\linewidth]{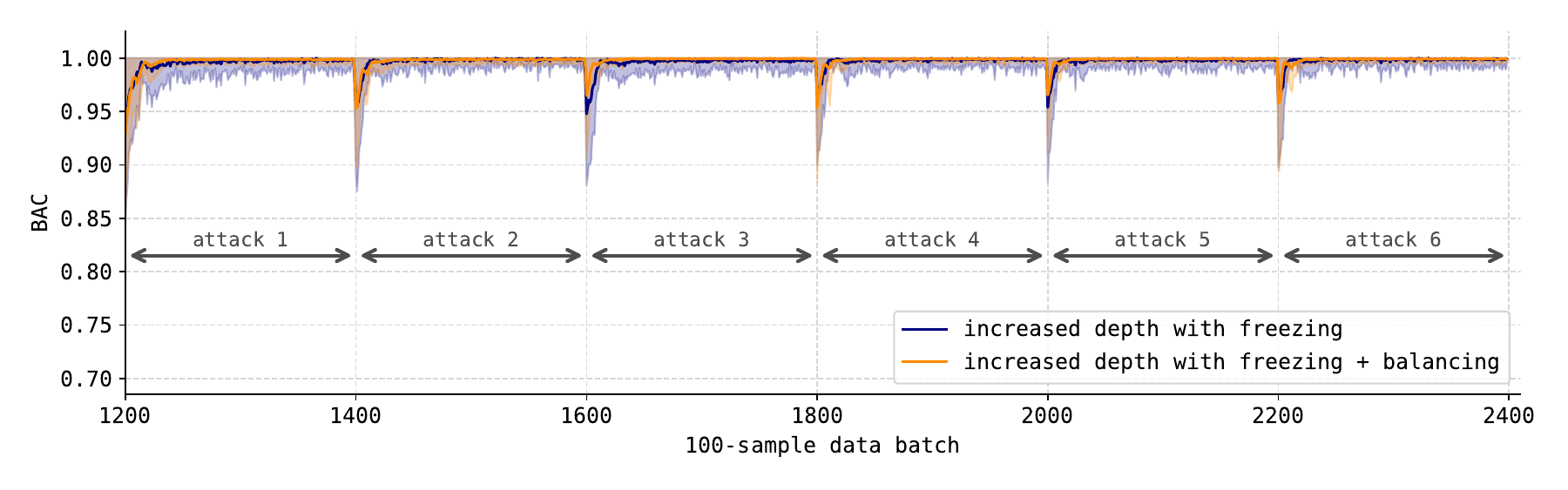}
    \caption{Balancing the dataset and increasing depth of the neural network with recurring attacks (only the recurring part--see x-axis). Average BAC with standard deviation. }
    \label{fig:balanced_and_increased_depth_recurring_with_freezing}
\end{figure*}

\begin{figure*}[t]
    \centering
    \includegraphics[width=\linewidth]{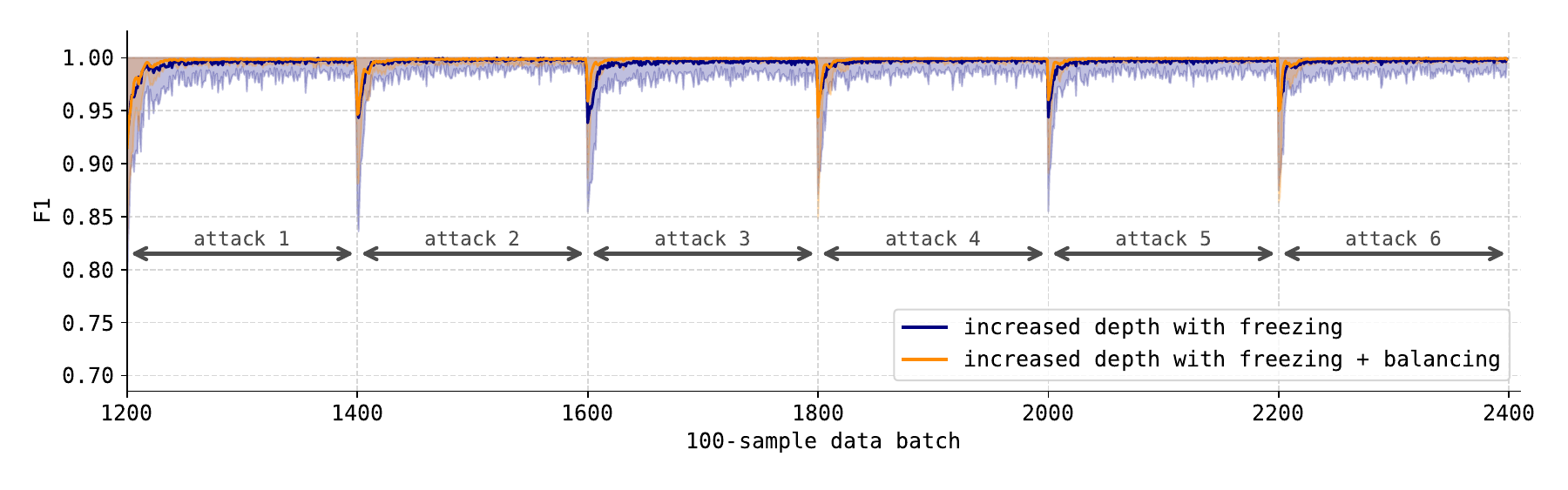}
    \caption{Balancing the dataset and increasing depth of the neural network with recurring attacks (only the recurring part--see x-axis). Average F1 with standard deviation. }
    \label{fig:f1_final_configuration}
\end{figure*}

\begin{figure*}[t]
    \centering
    \includegraphics[width=\linewidth]{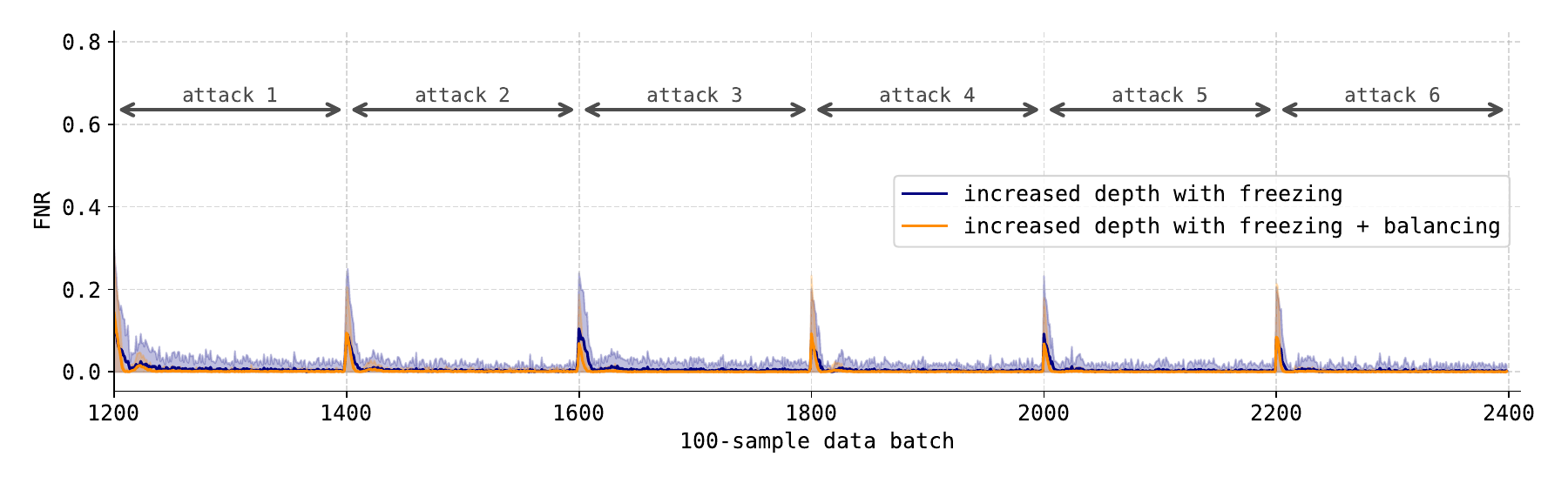}
    \caption{Balancing the dataset and increasing depth of the neural network with recurring attacks (only the recurring part--see x-axis). Average FNR with standard deviation. }
    \label{fig:fnr_final_configuration}
\end{figure*}

Finally, since the proposed and baseline mechanisms each bring certain benefits to attack detection, we evaluate whether their integration would yield meaningful advantages. Figure~\ref{fig:balanced_and_increased_depth_recurring} and Figure~\ref{fig:balanced_and_increased_depth_recurring_with_freezing} show the performance of their combination, compared to the baseline. Balancing the dataset in conjunction with increasing neural network depth further enhances the results. Primarily, we reduce the standard deviation, increasing the stability of the results. In the version without freezing hidden layers, the average standard deviation for each section improves by 73\%. In the second version it achieves 161\%. Additionally, the time to reach \ac{BAC} of 0.95 after an active attack change is improved by 37\% without freezing hidden layers, and by 57\% with freezing. This indicates that, in this final configuration, the accuracy drop is remarkably minimal and transient. For completeness, Figures~\ref{fig:f1_final_configuration} and \ref{fig:fnr_final_configuration} provide additional F1 and \ac{FNR} results for this final configuration, further confirming that the benefits of balancing persist as network depth increases and hidden-layer freezing is applied. In particular, \ac{FNR} provides an attack-specific perspective, measuring the proportion of attacks that remain undetected and thus highlighting the practical relevance of the proposed configuration. 

\section{Conclusions}
\label{sec:conclusions}
In this paper, we addressed the problem of recurring physical-layer attacks in dynamically evolving optical networks. We focused on the critical tradeoff between adapting to evolving attacks and retaining previously acquired knowledge of attack techniques over the long term. To this end, we proposed a~method for balancing the training data used to update attack detectors based on samples collected during attacks. A~threshold-based mechanism monitors the model's accuracy and triggers the balancing process when performance degrades, accelerating adaptation through a rehearsal strategy.

We demonstrated the effectiveness of the proposed approach through a~series of experiments using real data collected in a~laboratory environment. Our solution achieves performance comparable to a baseline from the literature that improves robustness by increasing the depth of the neural network. These results show that model performance can be improved simply by adjusting the proportions of data used for retraining, without modifying the network architecture. Finally, we combined both approaches, resulting in even more stable performance.

In the future, we plan to develop a hybrid system that combines the strengths of dynamic and static models. We also plan to extend our analysis to other \ac{ML} tasks in network optimization, including traffic prediction. 

\section*{Funding}
This work was supported by the statutory funds of the Department of Systems and Computer Networks, Wrocław University of Science and Technology, Poland, the European Commission through 5G-TACTIC (101127973), the Swedish Research Council (2023-05249), and VINNOVA through the Celtic-Next SUSTAINET-Advance project (2025-02987).

\section*{Acknowledgments} 
The authors thank Marco Schiano and Andrea Di Giglio for their contribution to collecting the dataset used in this work. We gratefully acknowledge Infinera (now part of Nokia) for providing the Groove G30 transponder.

\bibliography{bibliography}
  
\end{document}